%% file: JM_Cell_ICC_CR2.tex
\documentclass[10pt,letter,final]{IEEEtran}

\usepackage{amsthm,amssymb,graphicx,multirow,amsmath,color,amsfonts}%,ulem}
\usepackage[update,prepend]{epstopdf}
\usepackage[noadjust]{cite}
\usepackage[latin1]{inputenc}
\usepackage{tikz}
\usetikzlibrary{arrows,calc}		% Optional ticks libraries
\usepackage{bbm} % for \mathbbm{1}
\usepackage{pdfpages}
\usepackage{tabulary}
\usepackage{multirow}
\usepackage{comment}
\usepackage{dsfont}
\usepackage{pgf,tikz}
\usepackage{mathrsfs}
\usetikzlibrary{arrows}
\usepackage{bigints}
\usepackage{mathtools}
\newcommand{\bsloc}[1]{{\bf b}_{#1}}
\newcommand{\ueloc}[1]{{\bf u}_{#1}}

\newcommand{\lt}{\mathtt{LT}}

\newcommand{\pgfl}{\mathtt{PGFL}}
\newcommand{\cdf}{\mathtt{CDF}}
\newcommand{\pdf}{\mathtt{PDF}}

\newcommand\numberthis{\addtocounter{equation}{1}\tag{\theequation}}
\newcommand{\varn}[1]{\mathtt{Var}\big[#1\big]}

\newcommand{\numUE}{N}

\newcommand{\vCell}{{\cal V}}
\newcommand{\ccReg}{{\cal X}_C}
\newcommand{\ccAre}{{X}_C}

\include{notation}

\allowdisplaybreaks % Allows breaking of eqnarray over multiple pages (avoids unnecessary blanks in the document before eqnarray)

\begin{document}
\graphicspath{{./Figures/}}
\title{Johnson-Mehl Cell-based Analysis of UL Cellular Network with Coupled User and BS Locations}
\author{
Priyabrata Parida and Harpreet S. Dhillon \vspace{-0.75cm}
\thanks{The authors are with Wireless@VT, Department of ECE, Virginia Tech, Blacksburg, VA. Email: \{pparida, hdhillon\}@vt.edu. The support of the US National Science Foundation (Grant ECCS-1731711) is gratefully acknowledged.} % remove the date for conference drafts
}

\maketitle

\begin{abstract}
In this work, we analyze the performance of the uplink (UL) of a  cellular network where the base station (BS) locations follow a homogeneous Poisson point process (PPP), and the locations of users and BSs are spatially coupled.
In order to capture this coupling, we consider that users attached to a BS are located uniformly at random independently of each other in the Johnson-Mehl (JM) cell of that BS.
For this system model, we derive analytical expressions for the UL signal to interference ratio ($\sir)$ coverage probability and average spectral efficiency (SE) of a typical user in the network.
One of the key intermediate steps in our analysis is the approximate, but accurate, characterization of the area distribution of a typical JM cell. 
Another key intermediate step is the accurate statistical characterization of the point process formed by the interfering users that subsequently enables the coverage probability analysis. 
We present coverage probability and SE results for a typical user and study the interplay between different system parameters. 
\end{abstract}


\section{Introduction} \label{sec:intro}
In recent years, stochastic geometry has emerged as a powerful tool to analyze the performance of cellular networks.
The usual approach is to model the BS and the user locations as two independent Poisson point processes (PPP) and then analyze the performance of a randomly selected point from the user point process~\cite{AndBacJ2011, DhiGanJ2012, AndGupJ2016}.
The simplicity and tractability of this model have led to its wide-scale acceptance in the cellular research community.
Although it is a good first-order model, it suffers from one key shortcoming of not being able to capture inherent coupling in the user and BS locations, which results from the deployment of BSs in more populated areas.
Not surprisingly, this coupling in the locations has been accommodated in the simulation framework recommended by 3GPP (cf. \cite{3gppreportr12, 3gppreportr13}) for a few years now.
However, from stochastic geometry perspective, it was not until recently when new mathematical framework based on Poisson cluster process was introduced in~\cite{ManDP2016, AfshangD2016, SahaAD17B, SahaAD17a} to model and analyze the performance of this type of network.
Since the volume of the downlink (DL) traffic dominates the uplink (UL) traffic, the natural first step has always been to analyze the DL performance as has been carried out in the above-mentioned works.
However, due to a gradual shift from voice-only application to data-intensive applications, the asymmetry between the UL and DL traffic volume is marginalizing faster than ever before.
Therefore, it is important to analyze the UL performance of a cellular system while capturing the coupling between the BS and the user locations, which is the goal of this work.

{\em Motivation and related works:}
Since in UL the sources of interference are users, it is necessary to have a thorough understanding of the point processes formed by them.
% that is primarily dictated by the BS association policy.
Despite the traction that cellular  UL analysis using stochastic geometry has gained (cf. \cite{NovlanDA2013, ElSH2014, SinghZA14}), the understanding of the user point process is still evolving. 
In a few recent works (\cite{Haenggi2017, MonDas2017}), authors have analyzed the statistical properties of the user process for a scenario where the BSs are PPP distributed and the users are uniformly distributed within the Voronoi cell of each BS.
While these works take several important steps towards the accurate UL analysis, as mentioned earlier, their model does not capture the real-world location coupling, where the user density will be higher in certain region(s) (ideally in the proximity of a BS), which is a subset of the Voronoi cell of the BS.
{Building on the analysis presented in~\cite{AfshangDC16} for a clustered device-to-device network, the UL performance of a {closed-access} cellular system is presented in~\cite{TabHH2017}, where the coupling in the locations is taken into consideration.
To be specific, authors have modelled the user point process as a Mat\'{e}rn  cluster process (MCP), where their locations are uniformly distributed within circles centered at cluster head (BS) locations, which are PPP distributed.
As a result, two clusters of users can {\em overlap} resulting in some undesirable artifacts, such as higher density of users in the overlap region, which may not be realistic in a real-world setting. This particular limitation of an MCP model can be overcome by modelling user locations using Johnson-Mehl (JM) cells instead of just circular disks. As shown in Fig.~\ref{fig:ULDiagram}, this basically restricts the domain of the user locations within circular segments instead of circular disks.}
Although more realistic, the UL analysis using such a model is not straightforward due to the very fact that the statistical characterization of the point process formed by the interfering users is challenging.
In this paper, we overcome this challenge to enable the UL analysis of such a  network resulting in the following contributions.

{\em Contributions:} First, in order to capture the coupling between user and BS locations, we consider that users associated with a BS lie in a region of the Voronoi cell of the BS where each point of this region is within a predefined threshold distance from the BS.
This region, as we will discuss in detail in the sequel, is the JM cell of the BS.
Further, we present an approximate but accurate area distribution of a typical JM cell, which is not only helpful in characterizing the statistical properties of the user point process but also enables load (in terms of the number of users associated with a BS) based analysis for the UL.
Second, inspired by the recent work~\cite{Haenggi2017}, we present an accurate statistical characterization of the point process formed by the interfering user locations. 
Leveraging this new generative model, we provide tractable expression for the UL coverage probability of a typical user in the network.
In addition, using the cell area distribution, we also characterize the average spectral efficiency (SE) of a typical user.
{Further, using Monte Carlo simulations, we compare the UL coverage in the MCP based model used in \cite{TabHH2017} and the proposed JM cell-based model. 
The accuracy of all theoretical results is verified through extensive Monte Carlo simulations.}
%From this, we infer that at smaller threshold distances MCP based model overestimates the coverage probability while at larger threshold distances it underestimates the coverage.}

\section{System Model}\label{sec:SysMod}
{\em Network model:}
We consider a single tier network, where the locations of the BSs belong to the set $\Psi_b = \Phi_b \cup \{\nbo\}$, and the locations in $\Phi_b$ form a realization of a homogeneous PPP of density $\lambda_0$.
By virtue of Slyvniak's theorem, $\Psi_b$ is also a homogeneous PPP of density $\lambda_0$.
The location of the $j$-th BS is denoted by $\bsloc{j} \in \Psi_b$,
where index $j$ does not represent any ordering and $\bsloc{0} = \nbo = (0, 0)$ is at the origin.
In order to capture the coupling among the user and BS locations, we consider that users are uniformly distributed within the JM cell of each BS. 
While the JM cells can be described from the perspective of random nucleation and growth process~\cite{Moller1992}, we present a more intuitive definition for the JM cell associated with a typical BS.
Recall that the locations of the BSs can be used as seed points to form a Poisson-Voronoi tessellation (PVT) that completely covers $\R^2$ with convex sets known as Poisson Voronoi cells (PVC). 
Mathematically, the PVC of the typical BS at the origin ($0$-th BS) is given as 
\begin{align*}
\vCell_{\Psi_b}(\nbo) = \{\nbx \in \nbbR^2: \|\nbx\| \leq \|\nbx - \bsloc{j}\|, \forall \bsloc{j} \in \Psi_b\}.
 \numberthis
\end{align*}
For a given threshold radius $R_c$, JM cell of the typical BS is defined as the region of its PVC that is within a distance $R_c$ from its location, i.e. for the typical BS at the origin ($0$-th BS) we define its JM cell as ${\cal X}_C(\nbo, R_c, \Psi_b) =$
\begin{align*}
\{\nbx \in {\cal V}_{\Psi_b}(\nbo): \|\nbx\| \leq R_c\} = {\cal V}_{\Psi_b}(\nbo) \cap {\cal B}_{R_c}(\nbo), \numberthis
\label{eq:CCReg}
\end{align*}
where ${\cal B}_{R_c}(\nbo)$ denotes a circle of radius $R_c$ centred at $\nbo$.
We denote the area of the JM cell associated with the $j$-th BS (or with slight abuse of notation any typical BS) as $\ccAre(\lambda_0, R_c) = |{\cal X}_C(\bsloc{j}, R_c, \Psi_b)|$.
Let $\numUE_{Cj}$ be the number of users associated with the $j$-th BS.
We assume that $\numUE_{Cj}$ depends on the $j$-th JM cell area and follows a zero-truncated Poisson distribution with parameters $\lambda_u \ccAre(\lambda_0, R_c)$.
To be more precise, conditioned on the area of the $j$-th JM cell $X_C$, the probability mass function of $N_{Cj}$ is given as 
\begin{align*}
\dP{N_{Cj} = n| x_c} = \frac{\exp(-\lambda_u x_c) (\lambda_u x_c)^n}{n! (1-\exp(-\lambda_u x_c))}. \numberthis
\label{eq:UEPMF}
\end{align*}
One of the motivations behind consideration of the truncated Poisson distributions is to ensure that each BS in the network has at least one active user within its JM cell.
Consequently, this truncated Poisson distribution allows to model the user point process (to be defined shortly) as a Type-I process introduced in \cite{Haenggi2017}.
Note that $\lambda_u$ can be used to vary the load (the number of users per JM cell) in the network.

{We restrict our analysis to a narrow band single resource block system with bandwidth $B$.
Extension of the analysis to a system with multiple resource block is straightforward and is skipped in favour of simpler exposition.
Further, we assume that this resource block is shared among all the users associated to a BS in round robin manner. 
Note that at any given time, there is one active user in the JM cell of each BS.}
We present the performance analysis for a randomly selected user associated with the $0$-th BS that we term as the typical user in the network.
Let the point process formed by the locations of these active users be denoted as $\Psi_{\tt u}$, which is defined as
\begin{align*}
\Psi_{\tt u} = \{ U({\cal X}_C(\bsloc{j}, R_c, \Psi_b)): \forall \bsloc{j} \in \Psi_b\},
\end{align*}
where $U(B)$ denotes a uniformly distributed point in $B \subset \R^2$. From the construction, it is clear that the density of $\Psi_{\tt u}$ is $\lambda_0$.
Except the typical user attached to the $0$-th BS, rest of the users in the network are interfering users. 
Hence, the point processes formed by these interfering users is given as 
\begin{align*}
\Phi_{\tt u} = \{ U({\cal X}_C(\bsloc{j}, R_c, \Psi_b)) : \forall \bsloc{j} \in \Phi_b\}.
\end{align*}
We defer the discussion on the properties of the point process $\Phi_{\tt u}$ to Section~\ref{sec:CovAnalysis}.
%\chb{We find it pertinent to mention that although extension of this work to a system with multiple resource blocks is straightforward, it requires a more careful consideration of the point process formed by active interfering users in each cell. 
%This point process will be a thinned version of $\Phi_u$ where the thinning probability will depend on  $\lambda_u$ and the number of resource blocks.
%A detailed discussion on this will be accommodated in the future version of this work.}

\begin{figure}
  \centering
  \includegraphics[width=0.35\textwidth]{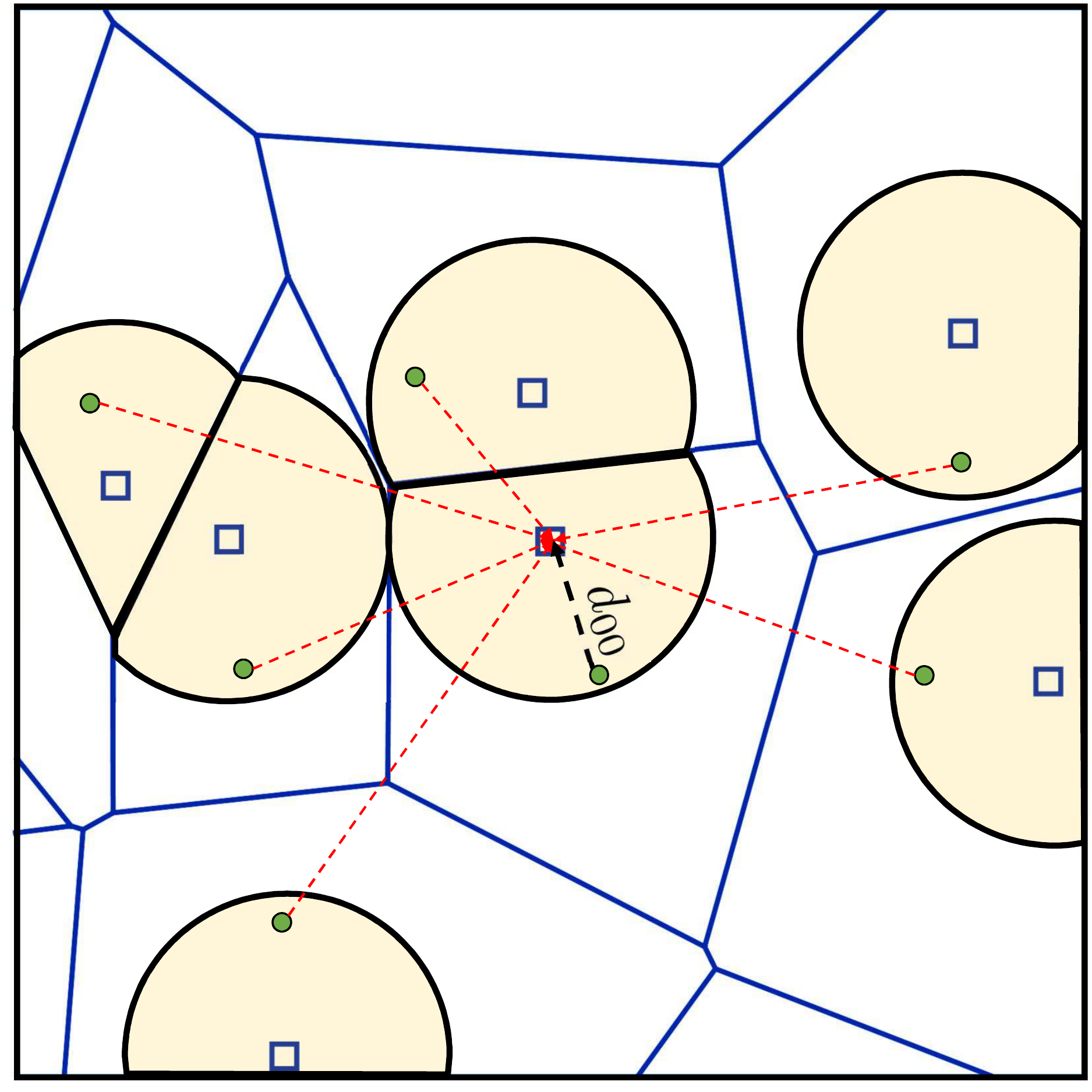}
  \caption{A representative network diagram for the UL with a single active user in each JM cell. The BS and user locations are denoted by squares and dots, respectively.}
  \label{fig:ULDiagram}
\end{figure}

Let the location of the active user attached to the $j$-th BS is denoted by $\ueloc{j}$.
Then, the distance between a user at $\ueloc{j}$ and a BS at $\bsloc{i}$ is given as $d_{ij} = \|\ueloc{j} - \bsloc{i}\|$.
{In order to characterize the coverage probability, the first step is the knowledge of the distribution of serving distance $D_{00}$ between the $0$-th BS and the typical user.
In case of a typical PVC, the distance distribution between the BS and a randomly located point in the PVC is approximated as Rayleigh distribution with scale parameter $(\sqrt{2 \pi \lambda_0 c_2})^{-1}$, where $c_2 = 5/4$ is an empirically obtained correction factor~\cite{YuMukIshYa2012, Haenggi2017}. 
Since, in our case, the user can not lie beyond the threshold radius $R_c$, it is reasonable to approximate the distribution $D_{00}$ to follow truncated Rayleigh distribution, which is given as}
\begin{align*}
F_{D_{00}} (d| R_c) = \frac{1 - \exp(- \pi c_2 \lambda_0 d^2)}{1 - \exp(- \pi c_2 \lambda_0 R_c^2)}. \numberthis
\label{eq:Djjk}
\end{align*}
At this point, we redefine $R_c$ in terms of normalized radius $\kappa$ as \mbox{\small
$
R_c = \kappa/\sqrt{\pi c_2 \lambda_0}, \quad \kappa \in [0, \infty)
$}. 
{In Sec.~\ref{sec:CovAnalysis}, $\kappa$ will be used to define the scale invariant pair correlation function (PCF) of $\Phi_{\tt u}$. This scale invariant property provides the flexibility to obtain PCF for $\lambda_0 = 1$, which can later be scaled to get the density function of $\Phi_{\tt u}$ for any value of $\lambda_0$.}
An illustrative diagram of the network is presented in Fig.~\ref{fig:ULDiagram}.

The channel gain between a BS and the typical user depends on small scale fading gain on the resource block, as well as the distance-dependent path loss. We assume that the small scale fading gain follows exponential distribution with mean $1$ and the path loss between two nodes at a distance $d$ is $d^{-\alpha}$, where $\alpha$ is the path loss exponent.
The consideration of shadowing is left for future work.
Under the above set of assumptions and in an interference limited scenario, the signal to interference ratio $(\sir)$ of the typical user associated with the $0$-th BS is given as 
\begin{equation}
{\sir}_{0} = h_{00} d_{00}^{-\alpha}(\sum\limits_{\ueloc{j} \in \Phi_{\tt u}} h_{0j} d_{0j}^{-\alpha})^{-1}, \numberthis
\label{eq:SinrULCI}
\end{equation}
where $h_{0j}$ is the small scale channel gain between the typical user and the $j$-th BS.

{\em Performance metrics:}
In this work, the system performance is evaluated using the following metrics.
\subsubsection{$\sir$ coverage probability} The $\sir$ coverage probability of the typical user for a target threshold $T$ is defined as
\begin{align*}
{\pc}(T, \kappa, \lambda_0) = &  \dP{\sir_0 > T}{}. \numberthis
\end{align*}

\subsubsection{Average user spectral efficiency} Considering round robin scheduling scheme, the average SE of a typical user is given as 
\begin{align}
\overline{\se}(\kappa, \lambda_0) = \dE{\frac{B}{N_{C0}} \log_2(1 + \sir_0)}{},
\label{eq:userSE}
\end{align}
where $B$ is the system bandwidth, and $N_{C0}$ is the number of users associated with the $0$-th BS.
In the following sections, we present our approach to obtain approximate but accurate theoretical expressions for the aforementioned quantities.

\section{Distribution of JM cell area of a typical BS}\label{sec:AreaDist}
The exact characterization of the distribution of a typical JM cell area is presented in \cite{Pineda2007}.
The expression of the probability density function $(\pdf)$ involves an infinite summation over multi-dimensional integrations. The order of integration depends on the number of neighbours that share the common boundary with the typical JM cell under consideration.
{Since the BS process is PPP, the number of common neighbours within a distance $2R_c$ from the BS location follows Poisson distribution, which is captured by the  infinite summation.}
%For instance, if the number of neighbours is $k$, then a $2k$-fold integration is necessary, and this needs to be sum over $\forall k \in \Z^+$. 
Although for smaller values of $R_c$ (equivalently $\kappa$), $\pdf$ using the expressions in \cite{Pineda2007} can be evaluated numerically, as $R_c$ increases the probability of having more number of neighbours increases, thereby increasing the complexity of the expression.
{This complexity limits its suitability to perform further analysis such as finding out the moments of the number of users in a typical JM cell.}
This motivates us to approximate the JM cell area distribution using a well-known distribution, which will be helpful not only in faster computational evaluation but also to get useful insights regarding different performance metrics by leveraging its known statistical properties.
In our approach, we first derive exact expressions for the first two moments of a typical JM cell area.
In the second step, using moment matching method, we approximate the cell area as a truncated beta distribution.
%{At the end of this section, the accuracy of the proposed distribution is validated through Monte-Carlo simulations with the help of statistical metrics such as Kulback-Leibler divergence (KLD) and Kolmogorov-Smirnov  distance (KSD). 

\subsubsection{Moments of JM cell area}
In the next Lemma, we derive the first two moments of the area of a typical JM cell.
\begin{lemma}\label{lem:M1M2CC}
For a given BS density $\lambda_0$ and threshold radius $R_c$, the mean area of a typical JM cell is given as
\begin{equation}\small{
m_{1, \ccAre}(\lambda_0, R_c) = \dE{\ccAre(\lambda_0, R_c)}{} = \frac{1 - \exp(- \pi \lambda_0 R_c^2)}{\lambda_0},
\label{eq:M1}}
\end{equation} 
and the second moment of the area is given by 
{\small
\begin{align*}
& m_{2, \ccAre}(\lambda_0, R_c) = \dE{\ccAre(\lambda_0, R_c)^2}{} \\
= & 2\pi \int_{r_1 = 0}^{R_c} \int_{r_2 = 0}^{R_c} \int_{u = 0}^{2 \pi} \exp\left(- \lambda_0 V(r_1, r_2, u) \right) {\rm d}u r_2{\rm d}r_2 r_1{\rm d}r_1, \numberthis
\label{eq:M2}
\end{align*}}
where \mbox{\small $V(r_1, r_2, u)$} is the area of union of two circles. The radii of these circles are $r_1$ and $r_2$, and the angular separation  between their centers with respect to origin is $u$.
Further,
{\small
\begin{align*}
V(r_1, r_2, u) = &   r_1^2 \left(\pi - v(r_1, r_2, u) + \frac{\sin(2v(r_1, r_2, u))}{2}\right) \\
& +  r_2^2 \left(\pi - w(r_1, r_2, u) + \frac{\sin(2w(r_1, r_2, u))}{2}\right), 
\label{eq:AoUC1C2}
\end{align*}
}%
where \mbox{\small $v(r_1, r_2, u) = \cos^{-1}\left(\frac{r_1 - r_2 \cos(u)}{\sqrt{r_1^2 + r_2^2 - 2r_1 r_2 \cos(u)}}\right)$}, and \mbox{\small $w(r_1, r_2, u) = \cos^{-1}\left(\frac{r_2 - r_1 \cos(u)}{\sqrt{r_1^2 + r_2^2 - 2r_1 r_2 \cos(u)}}\right)$}.
\end{lemma}
\begin{IEEEproof}
Please refer to Appendix~\ref{app:M1M2CC}.
\end{IEEEproof}

Using these expressions for the first two moments, the variance of a typical JM cell area is given as \mbox{\small
$
\varn{\ccAre(\lambda_0, R_c)} = m_{2, \ccAre}(\lambda_0, R_c) - m_{1, \ccAre}(\lambda_0, R_c)^2.
$}

\subsubsection{Approximate distribution of JM cell area}
Before arriving at the final distribution, some intuition on the type of distribution that provides an accurate approximation is necessary.
Recall that a typical JM cell can be expressed as the intersection of a PVC and a circle (ref.~\eqref{eq:CCReg}).
A PVC has two characteristic radii denoted as \mbox{\small $R_m$} and \mbox{\small $R_M$}~\cite{Calka2002}. While \mbox{\small $R_m$} corresponds to the radius of the largest circle that completely lies inside the Voronoi cell, \mbox{\small $R_M$} is the radius of the smallest circle that encircles the Voronoi cell. 
Note that \mbox{\small $R_m$} is half of the nearest neighbour distance of a PPP.
Since the nearest neighbour distance for a PPP follows Rayleigh distribution, the $\cdf$ of \mbox{\small $R_m$} is given as
\begin{equation}\small{
F_{R_m}(r_m) = 1 - \exp(-4 \pi \lambda_0 r_m^2).}
\label{eq:CDFRm}
\end{equation}
Now, we define an event {\small ${\cal E}_1 = \{R_c < R_m \} \equiv \{ {\cal B}_{R_c}(\nbo) \subset \vCell_{\Psi_b}(\nbo)\}$}.
The probability of \mbox{\small ${\cal E}_1$} is given as 
{\small
\begin{align*}
\dP{{\cal E}_1}{} = \dP{R_m > R_c} \stackrel{(a)}{=} \exp(-4 \pi \lambda_0 R_c^2) =  1 - \dP{{\cal E}_1^C}{}, \numberthis
\label{eq:PE1}
\end{align*}
}%
where $(a)$ follows from that $\cdf$ of \mbox{\small $R_m$} given in \eqref{eq:CDFRm}.
Observe that the $\pdf$ of \mbox{\small$\ccAre$} conditioned on \mbox{\small ${\cal E}_1$} is
{\small
\begin{align*}
f_{\ccAre}(x |{\cal E}_1) = \delta(\pi R_c^2), \numberthis
\label{eq:pdfAcE1}
\end{align*}
}%
where \mbox{\small $\delta(x)$} is the Dirac-delta function.
On the other hand, conditioned on the event \mbox{\small ${\cal E}_1^C$}, the $\pdf$ is continuous. Hence, the $\pdf$ of \mbox{\small $\ccAre$} is given as 
{\small
\begin{align*}
f_{\ccAre}(x) = f_{\ccAre}(x |{\cal E}_1) \dP{{\cal E}_1}{} + f_{\ccAre}(x |{\cal E}_1^C) (1 - \dP{{\cal E}_1}{}). \numberthis
\label{eq:pdfXc1}
\end{align*}
}%
The final part that remains to be determined in \eqref{eq:pdfXc1} is a suitable approximate distribution for \mbox{\small $f_{\ccAre}(x |{\cal E}_1^C)$}. 
In this work, we approximate \mbox{\small $f_{\ccAre}(x |{\cal E}_1^C)$} by generalized truncated beta distribution which is given as 
{\small
\begin{align*}
& f_{\ccAre}(x |{\cal E}_1^C) \approx  g(x; v, w, y, z, \alpha, \beta) \\
= & \frac{(x-y)^{\alpha-1} (z-x)^{\beta-1}}{\mathtt{B}(v, w, y, z; \alpha, \beta)}, \quad 0 \leq x  < \pi R_c^2, \numberthis
\label{eq:GenTrBeta}
\end{align*}
}%
where \mbox{\small $\alpha$} and \mbox{\small $\beta$} are shape parameters; the support of the untruncted beta distribution is \mbox{\small $[y, z]$} (since beta distribution has finite support); the support of the truncated beta distribution is \mbox{\small$[v, w]$}; and the normalization factor \mbox{\small $\mathtt{B}(v, w, y, z; \alpha, \beta)$} is {\small
\begin{align*}
\mathtt{B}(v, w, y, z; \alpha, \beta) = \int_{\mathfrak{v}}^{\mathfrak{w}} (x-y)^{\alpha-1} (z-x)^{\beta -1} {\rm d}x,
\end{align*}
}%
where \mbox{\small$\mathfrak{v} = \frac{v-y}{z-y}$} and \mbox{\small $\mathfrak{w} = \frac{w-y}{z-y}$}.
The choice of beta distribution is motivated by the fact that \mbox{\small $X_c$} has finite support \mbox{\small $[0, \pi R_c^2]$}.
Based on this, we set \mbox{\small$v = 0$ and $w = \pi R_c^2$} for the $\pdf$ presented in \eqref{eq:GenTrBeta}.
Another motivation behind choosing beta distribution is the presence of an additional shape parameter compared to conventional distributions such as Gamma or Weibull, which are parametrized by a single shape parameter.
In order to provide even more flexibility to closely match any arbitrary shape of the actual $\pdf$, we introduce truncation to the above distribution.
Hence, we set \mbox{\small$y = 0$} and \mbox{\small$z = 3/2 \pi R_c^2$}.
To obtain the shape parameters $\alpha$ and $\beta$, we follow the moment matching method, for which we need the mean and variance of \mbox{\small$\ccAre$} conditioned on \mbox{\small${\cal E}_1^C$}. In the following Lemma, we present aforementioned mean and variance.
\begin{lemma}\label{lem:CondCCAre}
Conditioned on \mbox{\small${\cal E}_1^C$}, the mean and variance of the area \mbox{\small$\ccAre$} is given as \mbox{\small $\dE{\ccAre | {\cal E}_1^C}{} (1-\dP{{\cal E}_1}) =$}
\begin{equation}\small{
\begin{aligned}
& \dE{\ccAre}{} - \dE{\ccAre | {\cal E}_1}{} \dP{{\cal E}_1}{}, \quad \varn{\ccAre | {\cal E}_1^C} (1-\dP{{\cal E}_1}) = \\
&   \varn{\ccAre} + 2 \dE{\ccAre|{\cal E}_1}{} \dP{{\cal E}_1} \dE{\ccAre|{\cal E}_1^C}{}\dP{{\cal E}_1^C} - \\
 & \dP{{\cal E}_1}(1-\dP{{\cal E}_1})(\dE{\ccAre|{\cal E}_1}{})^2 - \dP{{\cal E}_1^C}\dP{{\cal E}_1}(\dE{\ccAre|{\cal E}_1^C}{})^2.
\end{aligned}}
\end{equation}
\end{lemma}
\begin{IEEEproof}
The proof of this Lemma follows from law of total expectation and law of total variance that are given as \mbox{\small
$
\dE{\ccAre}{} = \dE{\ccAre | {\cal E}_1}{} \dP{{\cal E}_1}{} + \dE{\ccAre | {\cal E}_1^C}{} \dP{{\cal E}_1^C}, 
$}
and
{\small
\begin{align*}
\varn{\ccAre} = &  \varn{\ccAre|{\cal E}_1} \dP{{\cal E}_1} + \dP{{\cal E}_1}(1-\dP{{\cal E}_1})(\dE{\ccAre|{\cal E}_1}{})^2  \\
& + \varn{\ccAre|{\cal E}_1^C} \dP{{\cal E}_1^C} + \dP{{\cal E}_1^C}\dP{{\cal E}_1}(\dE{\ccAre|{\cal E}_1^C}{})^2 \\ 
& - 2 \dE{\ccAre|{\cal E}_1}{} \dP{{\cal E}_1} \dE{\ccAre|{\cal E}_1^C}{}\dP{{\cal E}_1^C}.
\end{align*}
}%
Rearranging the terms, and replacing \mbox{\small$\varn{\ccAre|{\cal E}_1} = 0$}, we obtain the expressions presented in the Lemma.
\end{IEEEproof}

The parameters \mbox{\small$\alpha, \beta$} for the $\pdf$ presented in \eqref{eq:GenTrBeta} can be obtained by solving the following system of equations
{\small
\begin{align*}
& \frac{\mathtt{B}(v, w, y, z; \alpha+1, \beta)}{\mathtt{B}(v, w, y, z; \alpha, \beta)} =  \dE{\ccAre | {\cal E}_1^C}{}, \\ 
& \frac{\mathtt{B}(v, w, y, z; \alpha+2, \beta)}{\mathtt{B}(v, w, y, z; \alpha, \beta)} - \dE{\ccAre | {\cal E}_1^C}{}^2  =  \varn{\ccAre|{\cal E}_1^C}. \numberthis
\label{eq:CCSysEq}
\end{align*}
}%
Now, the  approximate $\pdf$ for JM cell area is given as
{\small
\begin{align*}
f_{\ccAre}(x) =  \delta(\pi R_c^2) e^{-4 \pi \lambda_0 R_c^2} + f_{\ccAre}(x |{\cal E}_1^C) (1 -  e^{-4 \pi \lambda_0 R_c^2}), \numberthis
\label{eq:PDFCCArea}
\end{align*}
}%
where $f_{\ccAre}(x |{\cal E}_1^C)$ is given in \eqref{eq:GenTrBeta}.

\begin{figure}
  \centering
  \includegraphics[width=0.4\textwidth]{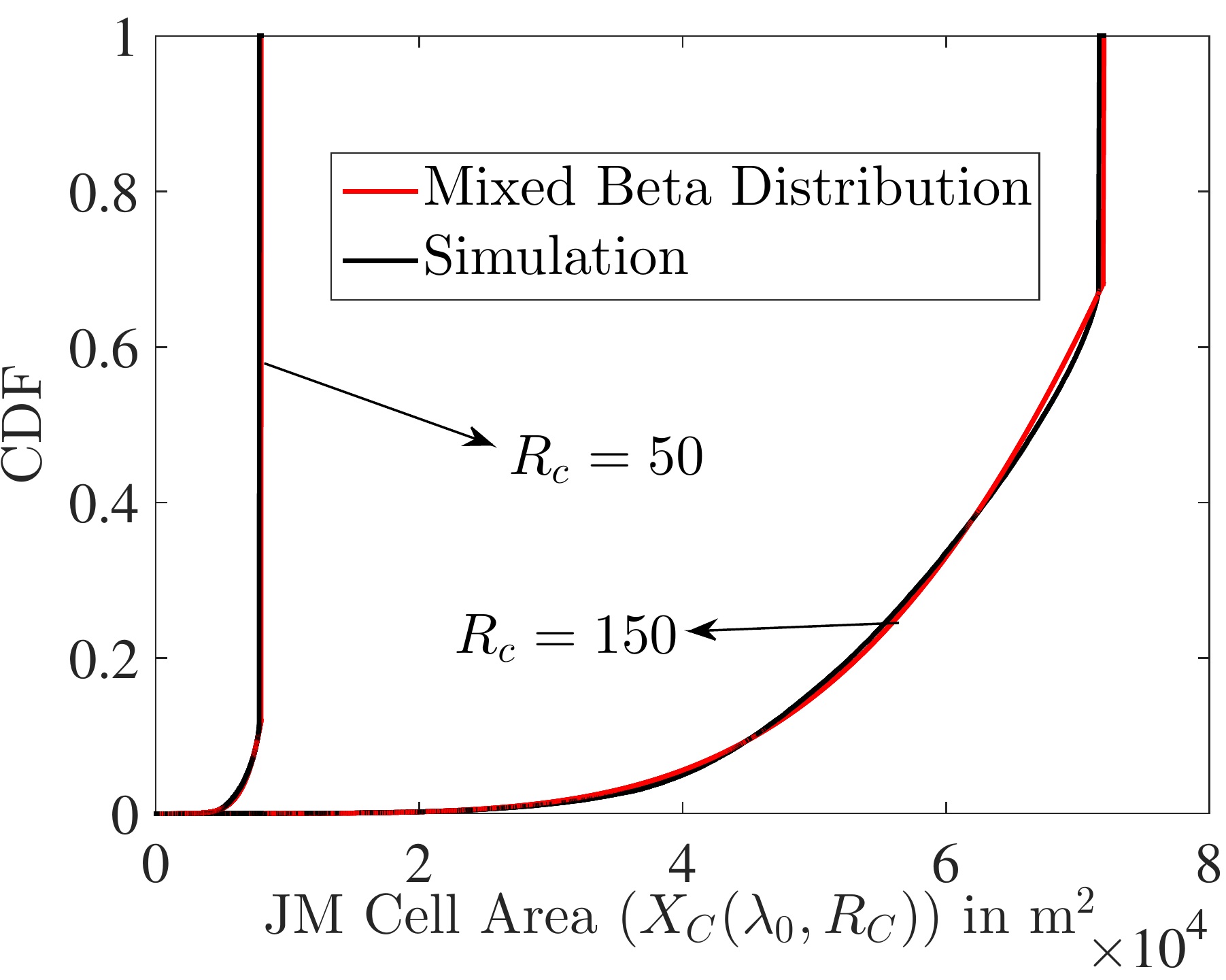}
  \caption{The $\cdf$ of a typical JM cell area for different $R_c$ (in m). $\lambda_0 = 4 \times 10^{-6}$ BS/m$^2$.}
  \label{fig:JMAreaDist}
\end{figure}

The approximate theoretical results are validated through Monte Carlo simulations.
In Table~\ref{tab:KSKL}, Kolmogorv-Smirnoff distance (KSD) and Kullback-Leibler divergence (KLD) are presented between the approximate and true (obtained through simulations) $\cdf$s ($\pdf$s) for different values of $R_c$.
{As observed from the table, the values of KSD and KLD are small for different $R_c$, which verifies the accuracy of the approximate distribution.}
In Fig.~\ref{fig:JMAreaDist}, we present the $\cdf$s of the area for visual verification purpose. 

\begin{table}[!htb]
\centering
\caption{Comparison between simulation and approximation $\cdf$s and $\pdf$s. $\lambda_0 = 4 \times 10^{-6}$.}
\begin{tabular}{|c|l|l|l|c|l|}
\hline
\multicolumn{1}{|c|}{$R_c | \kappa$}                           & \multicolumn{1}{c|}{$100 | 0.4$} & \multicolumn{1}{c|}{$200 | 0.8$} & \multicolumn{1}{c|}{$250 | 1$} & $300 | 1.2$     & $500 | 2$    \\ \hline
KSD  & 0.0230                   & 0.0238                   & 0.0123                   & \multicolumn{1}{l|}{0.0104} & 0.002  \\ \hline
KLD & 0.0125                   & 0.0095                   & 0.0055                   & 0.0032                      & 0.0007 \\ \hline
\end{tabular}
\label{tab:KSKL}
\end{table}

\section{$\sir$ Coverage and SE Analysis}\label{sec:CovAnalysis}
In Section~\ref{sec:SysMod}, we introduced the point process formed by the interfering user locations \mbox{\small$\Phi_{\tt u}$} without providing any details regarding its statistical properties. 
For coverage analysis the knowledge of the distribution of locations of users is essential.
The objective of this section is to characterize the statistical properties of \mbox{\small$\Phi_{\tt u}$} that is subsequently used to get the coverage probability expression.

\subsection{Density function of the interfering user point process}
{Conditioned on the typical BS at the origin, \mbox{\small$\Phi_{\tt u}$} is isotropic. In addition, since \mbox{\small$\Phi_{\tt u}$} is defined by excluding the typical user from \mbox{\small$\Psi_{\tt u}$}, it is non-homogeneous.
Now, our objective is to statistically characterize \mbox{\small$\Phi_{\tt u}$} conditioned on the $0$-th BS location.}
To achieve this objective, we follow the similar approach as presented in~\cite{Haenggi2017}. 
First, we determine the PCF, \mbox{\small$g(r)$}, of the interfering users with respect to (w.r.t.) the $0$-th BS.
Next, using the PCF, we approximate the point process as a non-homogeneous PPP. 

Note that in our case, the PCF \mbox{\small$g_{\lambda}(r, \kappa)$} is also a function of \mbox{\small$\kappa$}. By definition, \mbox{\small$g_{\lambda}(r, \kappa)$} presents the likelihood of finding a point of \mbox{\small$\Phi_{\tt u}$} at a distance $r$ from the $0$-th BS in a network with \mbox{\small$\lambda_0 = \lambda$} and threshold radius \mbox{\small$R_c = \kappa/\sqrt{\pi c_2 \lambda}$}. 
Further, the PCF is scale-invariant, i.e. \mbox{\small$g_{\lambda}(r, \kappa) = g_1(r \sqrt{\lambda}, \kappa)$}.
%where \mbox{\small$g_1(r, \kappa)$} represents the PCF for a network with \mbox{\small$\lambda_0 = 1$} and threshold radius \mbox{\small$\kappa/\sqrt{\pi c_2}$}.
Using the scale invariance property, in the following Lemma, we present the PCF of  \mbox{\small $\Phi_{\tt u}$} w.r.t. the origin for \mbox{\small $\lambda_0 = 1$}. 
\begin{lemma}\label{lem:PCFCC}
The PCF of interfering user locations w.r.t. the $0$-th BS is given as 
{\small
\begin{align*}
g_1(r, \kappa) \approx 1 - \exp\left(-2 \pi r^2 \dE{\ccAre(1, \kappa/\sqrt{\pi c_2})^{-1}}{}\right). \numberthis
\label{eq:PCFCC}
\end{align*}
}% 
\end{lemma}
\begin{IEEEproof}
Please refer to Appendix~\ref{app:PCFCC}.
\end{IEEEproof}

Using the above PCF, we approximate  \mbox{\small $\Phi_{\tt u}$} as a non-homogeneous PPP  such that for all  \mbox{\small $f : \R^2 \mapsto \R^+$}
{\small
\begin{align*}
& \mathbb {E}\bigg[ \sum _{x\in \Phi_{\tt u}} f(x) \bigg] = \mathbb {E} \bigg[\sum _{x\in \Phi_{\tt u}^{(\mathrm {PPP})} } f(x)\bigg]  \\
\implies   & \lambda_0 \int\limits_{\nbx \in \R^2} f(\nbx) g_1(\|\nbx\| \sqrt{\lambda_0}, \kappa) {\rm d}\nbx =  \int\limits_{\nbx \in \R^2} f(\nbx) \lambda_{\tt u}^{(\mathrm {PPP})}(\|\nbx\|, \kappa) {\rm d}\nbx,
\end{align*} 
}%
where the second step follows from the application of Campbell's theorem and replacing the intensity measure by the reduced second factorial moment measure~\cite[Chapter~8]{Haenggi2013}. Hence, the density of \mbox{\small$\Phi_{\tt u}$}, if approximated as a non-homogeneous PPP, is given as 
{\small
\begin{align*}
\lambda_{\tt u}^{(\mathrm {PPP})}(r, \kappa) =
 \lambda_0\left(1 - e^{- 2 \pi \lambda_0 r^2 \dE{X_C(1, \kappa/\sqrt{\pi c_2})^{-1}}{}}\right). \numberthis
\label{eq:IntUEDen}
\end{align*}
}%

\subsection{Coverage probability of a typical user}
To obtain the coverage probability, we first present the $\lt$ of aggregate interference in the following Lemma.
\begin{lemma}
The $\lt$ of aggregate interference at the $0$-th BS is given as
{\small
\begin{align*}
{\cal L}_{I_{\rm agg}}(s) = \exp\left(-2 \pi \int_{r=0}^{\infty} \frac{\lambda_{\tt u}^{(\mathrm{PPP})}(r, \kappa) r {\rm d}r}{1 + r^{\alpha} s^{-1}}\right).
\end{align*}
}%
\end{lemma}
\begin{IEEEproof}
As per the definition, the $\lt$ of aggregate interference is given as ${\cal L}_{I_{\rm agg}}(s) =$
{\small
\begin{align*}
& \dE{\exp(-s I_{\rm agg})}{\Phi_{\tt u}, \{h_{0j}\}}
= \dE{\prod_{\ueloc{j} \in \Phi_{\tt u}} \dE{\exp(-s h_{0j} d_{0j}^{-\alpha})}{h_{0j}}}{} \\
= &  \dE{\prod_{\ueloc{j} \in \Phi_{\tt u}} \frac{1}{1 + s d_{0j}^{-\alpha}}}{} 
= \exp\left(-2 \pi \int_{r=0}^{\infty} \frac{\lambda_{\tt u}^{(\mathrm{PPP})}(r, \kappa) r {\rm d}r}{1 + r^{\alpha} s^{-1}}\right),
\end{align*}
}%
where the last step follows from the application of $\pgfl$ of PPP.
\end{IEEEproof}

Now using the $\lt$ of interference, in the following proposition, we present the coverage probability of a randomly selected user associated with the typical BS. 
\begin{prop}
For a target $\sir$ threshold $T$, the UL coverage probability of a typical user is given as 
{\small
\begin{align*}
\pc(T, \kappa, \lambda_0) = \int\limits_{r = 0}^{R_c} {\cal L}_{I_{\tt agg}}(r^{\alpha} T) f_{D_{00}}(r|R_c) {\rm d}r. \numberthis
\label{eq:Pc}
\end{align*}
}%
\end{prop}
\begin{IEEEproof}
The coverage probability for the typical user is defined as 
{\small 
\begin{align*}
\dP{\sir_0 > T} = \dP{h_{00} > d_{00}^{\alpha}T I_{agg}} 
= \dE{e^{-d_{00}^{\alpha}T I_{\rm agg}}}{D_{00}},
\end{align*}
}%
where \eqref{eq:Pc} follows from deconditioning w.r.t. $D_{00}$.
\end{IEEEproof}

Using \eqref{eq:Pc} and the area distribution of a typical JM cell, in the following proposition, we present the average achievable SE of a typical user.
\begin{prop}
The average SE of a typical user is given as 
{\small
\begin{align*}
\overline{\se}(\kappa, \lambda_0, \lambda_u) = B\dE{N_{C0}^{-1}}{} \int\limits_{t=0}^{\infty} \pc(2^t-1, \kappa, \lambda_0) {\rm d}t, \numberthis
\label{eq:userSEAprx}
\end{align*}
}%
where 
$
\dE{N_{C0}^{-1}}{} =  \int\limits_{x_c=0}^{\pi R_c^2} \sum_{n=1}^{\infty} \frac{\dP{N_{C0}=n|x_c}}{n}  f_{X_c}(x_c) {\rm d}x_c.
$
\end{prop}

\begin{IEEEproof}
Assuming independence between $N_{C0}$ and $\sir_0$, \eqref{eq:userSE} can be approximately expressed as
{\small
\begin{align*}
\overline{\se}(\kappa, \lambda_0) = B\dE{N_{C0}^{-1}}{} \dE{\log_2(1 + \sir_0)}{}.
\end{align*}
}%
The expression in \eqref{eq:userSEAprx} follows from the fact that for a positive random variable $X$, $\dE{X}{} = \int_{t=0}^{\infty} \dP{X > t} {\rm d}t$.
\end{IEEEproof}

\section{Results}

In this section, we verify the accuracy of the approximate theoretical expressions using Monte Carlo simulations. 
%Further, we study the effect of different system parameters on the $\sinr$ coverage probability, average user SE. 
We consider the BS density $\lambda_0 = 4 \times 10^{-6}$ BS/m$^2$, and path loss exponent $\alpha = 3.7$. 
The system bandwidth $B$ is taken to be 1 Hz to focus on user SE.

\begin{figure}[!htb]
  \centering
  \includegraphics[width=0.44\textwidth]{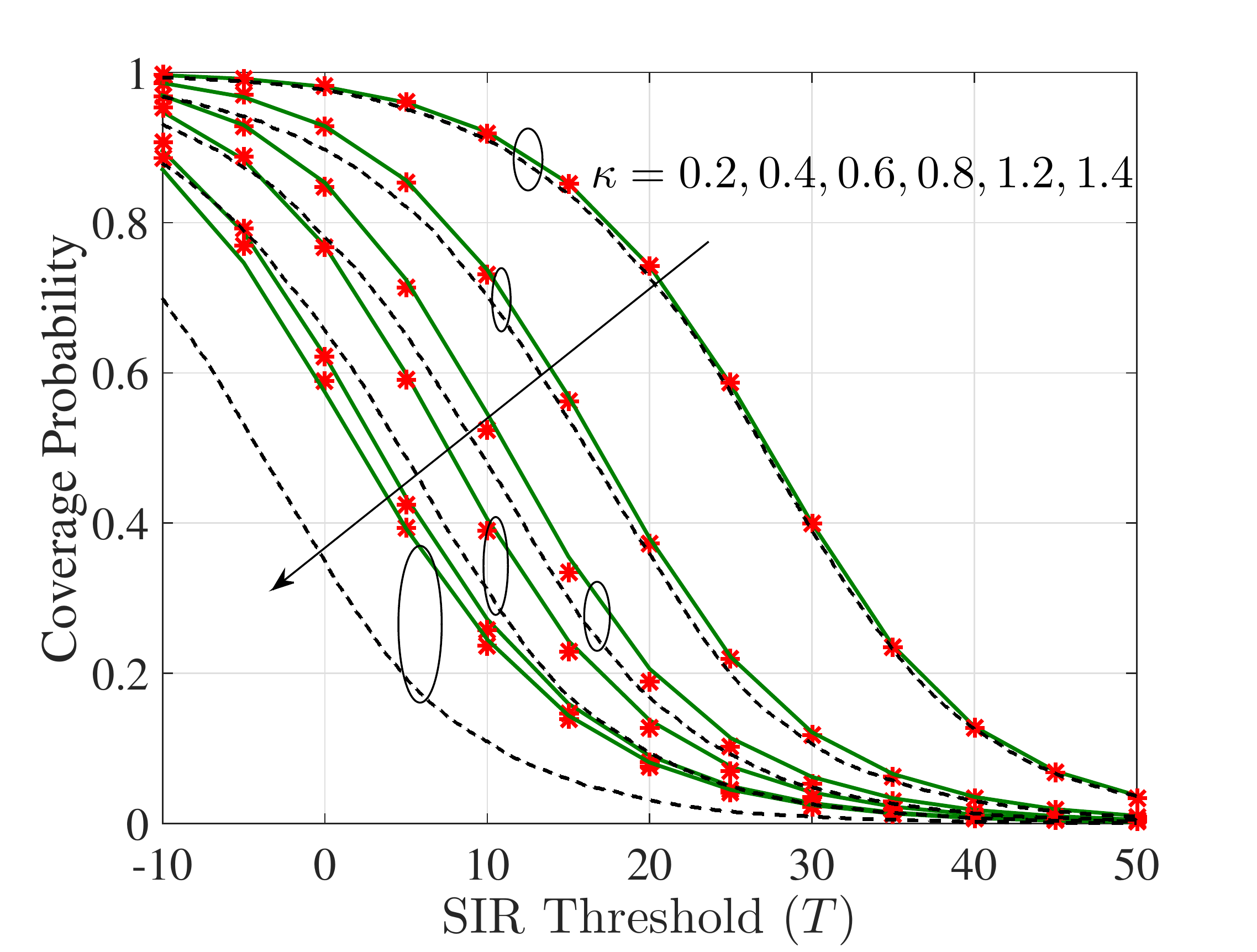}
  \caption{The $\sir$ coverage probability of a typical user. Markers and solid lines represent the simulation and theoretical results, respectively, based on the proposed model. The dashed line represent the coverage probability of a typical user based on MCP model. $\lambda_0 = 4 \times 10^{-6}$ BS/m$^2$, $\lambda_u = 200 \lambda_0$ users/m$^2$.}
  \label{fig:PcKappa}
\end{figure}

The $\sir$ coverage probability of a typical user in UL is presented in Fig.~\ref{fig:PcKappa} for different values of $\kappa$ (equivalently $R_c$).
As observed from the figure, with increasing $\kappa$ reduction in average serving distance results in coverage probability degradation. 
{Further, in order to highlight the usefulness of the proposed model, through Monte Carlo simulations, we present the coverage probability (dashed black lines) of a typical user using MCP based model~\cite{TabHH2017}.
As observed from the figure, with respect to the proposed model, the MCP-based model underestimates the coverage probability.}

\begin{figure}[!htb]
  \centering
  \includegraphics[width=0.44\textwidth]{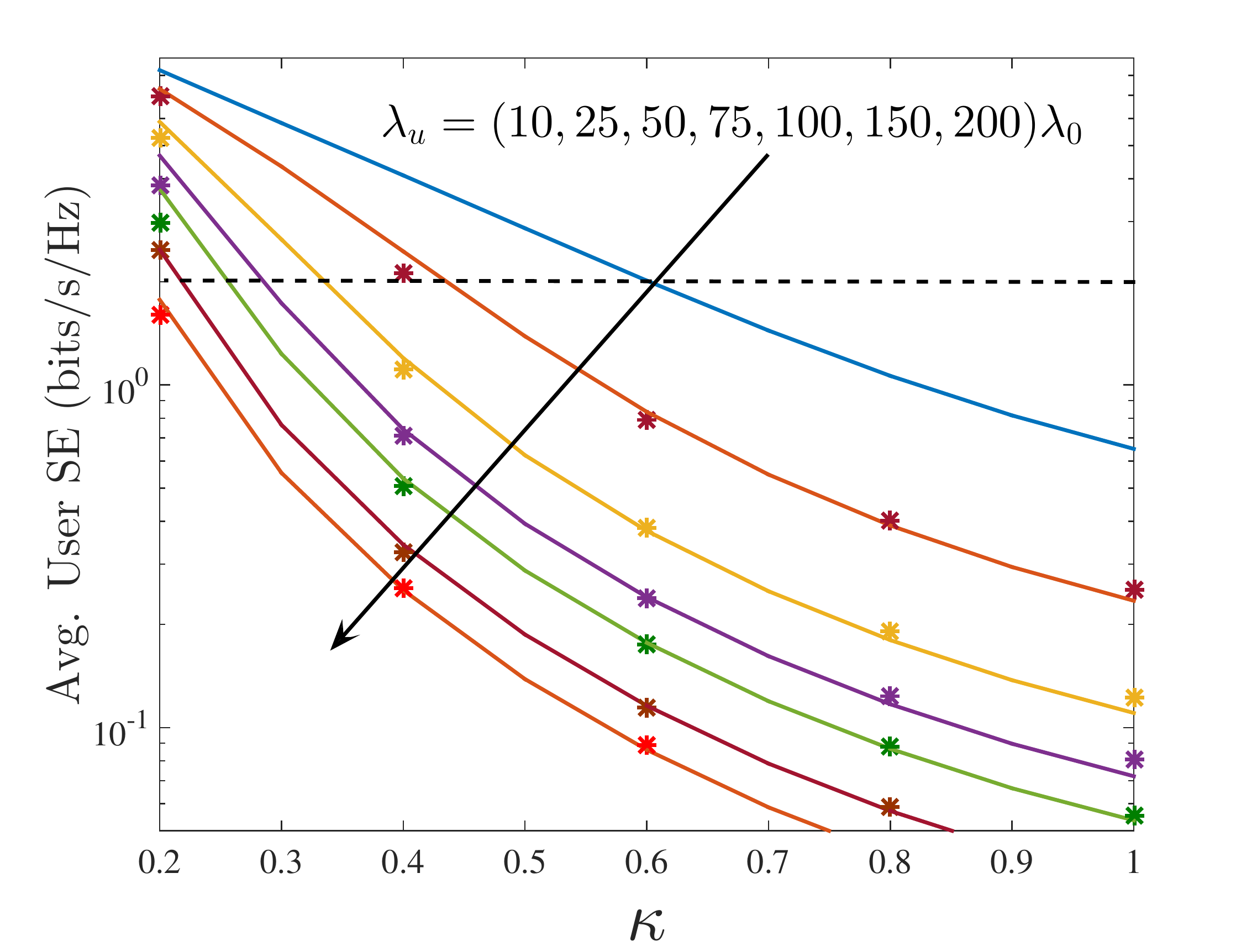}
  \caption{The average spectral efficiency of a typical user.  Markers represent simulation results. $\kappa = R_c\sqrt{\pi c_2 \lambda_0}, \lambda_0 = 4 \times 10^{-6}$ BS/m$^2$.}
  \label{fig:USEvsKappa}
\end{figure}

The average SE of the typical user is presented in Fig.~\ref{fig:USEvsKappa}.
As observed from the figure, the average SE decreases with increasing $\kappa$. This is justified by the fact that with increasing $\kappa$ the coverage probability reduces.
Further, due to increasing number of users, the typical user is assigned the resource block less frequently.
From this figure, we also gain insights regarding the achievable average user SE for given $\lambda_u, \kappa$. For example, a network with $\lambda_u = 200 \lambda_0$ users/m$^2$ and $\kappa \geq 0.2$, cannot support an average user SE of $2$ bits/s/Hz.
In both the figures, the theoretical results closely match with the simulation results, which verifies the accuracy of the theoretical expressions.

\section{Conclusion}
In this work, we proposed a new model to analyze the UL performance of a cellular network considering coupling between BS and user locations. 
This coupling is captured by modeling the users to be uniformly distributed in the JM cells of each BS.
The first important result of this work is the approximate area distribution of a typical JM cell that can be used to model the load distribution (in terms of the number of users) in a typical BS.
The second important outcome of this work is the statistical characterization of the point process formed by the interfering user locations that is later used to derive the UL coverage probability for a typical user.
The accuracy of the approximate theoretical expressions is verified through Monte Carlo simulations using two key metrics: coverage probability and average user spectral efficiency.
%One important conclusion is that with respect to the proposed model, which is more realistic, the existing MCP based model either overestimates or underestimates the coverage probability of a typical user.
{The proposed model and corresponding analyses in this work can be extended in many directions to obtain important system design guidelines, some of which will be accommodated in the future version of this work.}
 
\appendix
\subsection{Proof of Lemma~\ref{lem:M1M2CC}}\label{app:M1M2CC}
%A general expression to obtain the $n$-th moment of the area a typical PVC is presented in~\cite[Theorem 3.1]{alishahi2008}.
%We use the similar approach to arrive at the expressions presented in the Lemma.
Note that a typical JM cell can be represented as the region of intersection between a PVC and a circle.
Hence, the mean area of a typical JM cell can be expressed as \mbox{\small $\dE{|{\cal X}_C(\nbo, R_c, \Psi_b)|}{} =$}
{\small
\begin{align*}
 & \dE{\int\limits_{\nbx \in \nbbR^2} \mathbf{1}_{(\nbx \in \vCell_{\Psi_b}(\nbo) \cap {\cal B}_{R_c} (\nbo))} {\rm d}\nbx}{} =  \int\limits_{\nbx \in \nbbR^2 \cap {\cal B}_{R_c} (\nbo)} {\rm d}\nbx \dE{\mathbf{1}_{(\nbx \in \vCell_{\Psi_b}(\nbo))}}{} \\
& \stackrel{(a)}{=} \int\limits_{\nbx \in \nbbR^2 \cap {\cal B}_{R_c} (\nbo)} \exp(-\pi \lambda_0 \|\nbx\|^2) {\rm d}\nbx = 2 \pi \int\limits_{r=0}^{R_c} \exp(-\pi \lambda_0 r^2) r {\rm d}r,
\end{align*}
}%
where $(a)$ follows from that fact that a point located at a distance $\|\nbx\|$ from the origin belongs to $\vCell_{\Psi_b}(\nbo)$, if there are no other BSs in ${\cal B}_{\|\nbx\|}(\nbx)$.
Solving the final integral gives us the expression for the mean in \eqref{eq:M1}.
Similarly, the second moment of the area of a JM cell can be expressed as \mbox{\small $\dE{|{\cal X}_C(\nbo, R_c, \Psi_b)|^2}{} =$}
{\small
\begin{align*}
&  \dE{\int\limits_{\nbx \in \nbbR^2} \mathbf{1}_{(\nbx \in \vCell_{\Psi_b}(\nbo) \cap {\cal B}_{R_c} (\nbo))} {\rm d}\nbx \int\limits_{\nby \in \nbbR^2} \mathbf{1}_{(\nby \in \vCell_{\Psi_b}(\nbo) \cap {\cal B}_{R_c} (\nbo))} {\rm d}\nby}{} \\
= & \int\limits_{\nbx \in \nbbR^2} \int\limits_{\nby \in \nbbR^2} \dE{\mathbf{1}_{(\nbx \in \vCell_{\Psi_b}(\nbo)\cap {\cal B}_{R_c} (\nbo), \nby \in \vCell_{\Psi_b}(\nbo) \cap {\cal B}_{R_c} (\nbo))}}{} {\rm d}\nby  {\rm d}\nbx \\
\stackrel{(b)}{=} & \int\limits_{(\nbx, \nby) \in \nbbR^2 \cap {\cal B}_{R_c} (\nbo) \times \nbbR^2 \cap {\cal B}_{R_c} (\nbo)} e^{-\lambda_0 |{\cal B}_{\|\nbx\|}(\nbx) \cup {\cal B}_{\|\nby\|}(\nby)|} {\rm d}\nbx {\rm d}\nby \\
= & 2 \pi \int\limits_{r_1=0}^{R_c} \int\limits_{r_2=0}^{R_c} \int\limits_{u=0}^{2 \pi} e^{-\lambda_0 V(r_1, r_2, u)} {\rm d}u r_2{\rm d}r_2 r_1{\rm d}r_1,
\end{align*}
}%
where $(b)$ follows from the fact that if points $\nbx$ and $\nby$ belong to $\vCell_{\Psi_b}(\nbo)$, then there are no other BSs in the region ${\cal B}_{\|\nbx\|}(\nbx) \cup {\cal B}_{\|\nby\|}(\nby)$,  and the last step follows from changing the integration limits from Cartesian to polar coordinates.\hfill 
\IEEEQED

\subsection{Proof of Lemma~\ref{lem:PCFCC}}\label{app:PCFCC}
One approach to deriving \mbox{\small $g_1(r, \kappa)$} is to first determine the Ripley's K-function \mbox{\small $K_1(r, \kappa)$} and then exploit the following relationship: \mbox{\small
$
g_1(r, \kappa) = \frac{{\rm d}K_1(r, \kappa)/{\rm d}r}{2 \pi r}.
$}
Observe that the points in \mbox{\small $\Phi_{\tt u}$} are likely to exhibit repulsion w.r.t. \mbox{\small $\nbo$} as these users (points) being interfering users do not belong to \mbox{\small ${\cal X}_C(\nbo, R_c, \Psi_b)$}.
Since $g_1(r, \kappa) \rightarrow 1$ for \mbox{\small $r \gg 0$},
our main interest lies in characterizing \mbox{\small $g_1(r, \kappa)$} for small \mbox{\small $r$}, which we do next.

Recall that for a point process \mbox{\small $\Phi$} of density \mbox{\small$\lambda$}, the Ripley's K-function is defined as \mbox{\small$K_\lambda(r) = \dE{N_{\Phi}({\cal B}_r(\nbo))}{}/\lambda$}~\cite{Haenggi2013}, where \mbox{\small$N_{\Phi}({\cal B}_r(\nbo))$} denotes the number of points of \mbox{\small$\Phi$} that lie in \mbox{\small${\cal B}_r(\nbo)$}.
In this case, the K-function is given as \mbox{\small 
$
K_1(r, \kappa) = \dE{N_{\Phi_{\tt u}}\left(\cup_{\nbx \in \Phi_b}({\cal B}_r(\nbo) \cap \ccReg(\nbx, \kappa/\sqrt{\pi c_2}, \Psi_b))\right)}{}.
$}
As \mbox{\small $r \rightarrow 0$}, we can write
{\small
\begin{align*}
K_1(r, \kappa) \sim \dE{N_{\Phi_{\tt u}}\left({\cal B}_r(\nbo) \cap \ccReg(\nby, \kappa/\sqrt{\pi c_2}, \Psi_b)\right)}{}, \numberthis
\label{eq:K1_1}
\end{align*}
}%
where \mbox{\small $\nby$} is the nearest BS to the typical BS at $\nbo$.
Without loss of generality, we assume that \mbox{\small $\nby = (\|\nby\|, 0)$}.
As per our construction of \mbox{\small $\Phi_{\tt u}$}, 
we are concerned with only one uniformly distributed point in \mbox{\small $\ccReg(\nby, \kappa/\sqrt{\pi c_2}, \Psi_b)$} lying in the region \mbox{\small ${\cal B}_r(\nbo) \cap \ccReg(\nby, \kappa/\sqrt{\pi c_2}, \Psi_b)$}.
Hence, we write \eqref{eq:K1_1} as \mbox{\small{ $K_1(r, \kappa) \sim$}}
{\small
\begin{align*}
& \dE{\frac{|{\cal B}_r(\nbo) \cap \ccReg(\nby, \kappa/\sqrt{\pi c_2}, \Psi_b)|}{|\ccReg(\nby, \kappa/\sqrt{\pi c_2}, \Psi_b)|}}{} = \dE{\frac{S_C(r_m, r, \kappa)}{\ccAre(1, \kappa/\sqrt{\pi c_2})}}{} \\
\approx & \dE{S_C(r_m, r, \kappa)}{R_m} \dE{\ccAre(1, \kappa/\sqrt{\pi c_2})^{-1}}{},
\end{align*}
}%
where \mbox{\small $S_C(r_m, r, \kappa)$} denotes the area of the region \mbox{\small ${\cal B}_r(\nbo) \cap {\cal B}_{R_c}(\nby) \cap \left((\nbbR - r_m)^+ \times \nbbR\right)$}, and the last approximation follows from independence assumption between \mbox{\small $S_C(r_m, r, \kappa)$} and \mbox{\small $X_C(1, \kappa/\sqrt{\pi c_2})$}.
Now, using the result presented in Appendix~\ref{app:ExpArea}, we write
{\small
\begin{equation}
\dE{S_C(r_m, r, \kappa)}{R_m} \sim
\begin{cases}
\frac{\pi^2 r^4}{2}, & R_c > r, r \rightarrow 0 \\
\pi^2 R_c^2 r^2 - \frac{\pi^2 R_c^4}{2}, &  R_c \leq r, r \rightarrow 0,
\end{cases}
\label{eq:ExpArea}
\end{equation}
}%
where \mbox{\small $R_c = \kappa/\sqrt{\pi c_2}$.}
The first inverse moment of \mbox{\small $\ccAre(1, \kappa/\sqrt{\pi c_2})$} can be evaluated numerically using the approximated distribution presented in Section~\ref{sec:AreaDist}. 
Now, the K-function is given as \mbox{\small $K_1(r, \kappa) \approx $}
{\small
\begin{align*}
\begin{dcases}
\pi^2 r^4/2 \dE{X_C(1, \kappa/\sqrt{\pi c_2})^{-1}}{} & R_c > r, r \rightarrow 0 \\
(\pi^2 R_c^2 r^2 - \pi^2 R_c^4/2) \dE{X_C(1, \kappa/\sqrt{\pi c_2})^{-1}}{} & R_c \leq r, r \rightarrow 0,
\end{dcases}
\end{align*}
}%
and the PCF is given as \mbox{\small $g_1(r, \kappa) =$}
{\small
\begin{align*}
\frac{{\rm d}K_1(r,\kappa)/{\rm d}r}{2 \pi r} \approx
\begin{dcases}
\pi r^2 \dE{X_C(1, \kappa/\sqrt{\pi c_2})^{-1}}{} &  R_c > r, r \rightarrow 0 \\
\pi R_c^2 \dE{X_C(1, \kappa/\sqrt{\pi c_2})^{-1}}{} & R_c \leq r, r \rightarrow 0.
\end{dcases}
\end{align*}
}%
Note that when \mbox{\small $R_c \rightarrow 0$, $\dE{X_c(1, \kappa/\sqrt{\pi c_2})^{-1}}{} \sim \frac{1}{\pi R_c^2}$}.
Hence, as expected, \mbox{\small $g_1(r, \kappa) \sim 1$} as \mbox{\small $R_c \rightarrow 0$}. 
Using the asymptotic result that \mbox{\small $1 - \exp(-u) \sim u$} as \mbox{\small $u \rightarrow 0$}, we write 
{\small
\begin{align*}
g_1(r, \kappa) \approx 
\begin{dcases}
1 -e^{- \pi r^2 \dE{X_C(1, \kappa/\sqrt{\pi c_2})^{-1}}{}} & r < R_c, r \rightarrow 0 \\
1 & r \geq R_c , r \rightarrow 0.
\end{dcases}
\end{align*}
}%
As per the simulation based observation mentioned in \cite{Haenggi2017}, due to the condition \mbox{\small $r \rightarrow 0$}, the Voronoi cell \mbox{\small ${\cal V}_{\Psi_b}(\nby)$} is skewed whose area is likely to be half of area of a typical Voronoi cell. 
Similar argument can be made for the JM cell shape and area.
Hence, a factor of 2 needs to be introduced for the first condition.
Using this fact, for any value of $r$, a reasonably good approximation for the PCF is \mbox{\small
$
g_1(r, \kappa) \approx 1 -\exp(- 2 \pi r^2 \dE{X_C(1, \kappa/\sqrt{\pi c_2})^{-1}}{}). 
$}
%since \mbox{\small $g_1^{\rm CC}(r, \kappa) \rightarrow 1$} as \mbox{\small $r >> 0$}.
This completes the proof of the Lemma.\hfill
\IEEEQED

\subsection{Proof of \eqref{eq:ExpArea}}\label{app:ExpArea}
Depending on the value of \mbox{\small $R_c$} and \mbox{\small $r$} we have following two cases of interest:\newline
{\bf Case~1:} \mbox{\small $r < R_c$}: The result for this case is obtained from \cite[Lemma~2]{Haenggi2017}, and is given as
{\small
\begin{align*}
\dE{S_C(r_m, r, \kappa)}{R_m} \sim \frac{\pi^2 r^4}{2} , \quad r \rightarrow 0. 
\end{align*}
}% 
{\bf Case~2:} \mbox{\small $r \geq R_c$}: In this case, the area of the region \mbox{\small ${\cal B}_r(\nbo) \cap \ccReg(\nby, \kappa/\sqrt{\pi c_2}, \Psi_b)$} is given as
{\small
\begin{align*}
S_C(r_m, r, \kappa) =
\begin{dcases}
r^2 \left(u - \frac{\sin{2u}}{2}\right) + R_c^2 \left(v - \frac{\sin{2v}}{2}\right) & \\
\quad - (w R_c^2 - r_m \sqrt{R_c^2 - r_m^2}),  & R_c \geq r_m\\
r^2 u - \frac{r^2}{2}\sin{2u} + R_c^2 v - \frac{R_c^2}{2}\sin{2v}, & R_c < r_m 
\end{dcases}
\end{align*}
}%
where {\small $R_c = \kappa/\sqrt{\pi c_2}$, $u = \cos^{-1}\left(\frac{r^2 + 4 r_m^2 -R_c^2}{4 r r_m}\right)$, $v = \cos^{-1}\left(\frac{R_c^2 + 4 r_m^2 - r^2}{4 R_c r_m}\right)$, and $w = \cos^{-1}\left(\frac{r_m}{R_c}\right)$.}
%We can replace $R_c = a r$ where $a \in [0, 1)$.
Averaging over the random variable $R_m$, we get $\dE{S(r_m, r, \kappa)}{} =$
{\small
\begin{align*}
\pi R_c^2  \int\limits_{0}^{(r-R_c)/2} f_{R_m}(r_m) {\rm d}r_m + \int\limits_{(r-R_c)/2}^{(r+Rc)/2} S_C(r_m, r, \kappa) f_{R_m}(r_m) {\rm d}r_m,
\end{align*}
}%
where we have used the fact that for $\mbox{\small $r > 2 r_m + R_c$}$, \mbox{\small $S_C(r_m, r, \kappa) = \pi R_c^2$}.
Further, note that for \mbox{\small$2 r_m > r + R_c$,  $S_C(r_m, r, \kappa) = 0$}. Hence, the upper limit in the second integration is introduced to consider the values of \mbox{\small $R_m$} for which \mbox{\small $S_C(r_m, r, \kappa) \neq 0$.}
In addition, we use the asymptotic approximation \mbox{\small $f_{R_m}(r_m) =  8 \pi r_m \exp(-4 \pi r_m^2) \sim 8 \pi r_m (1 - 4 \pi r_m^2)$}, as \mbox{\small $r_m \rightarrow 0$.}
After performing the integration, we obtain \mbox{\small $\dE{S(r_m, r, \kappa)}{} \sim $}
{\small
\begin{align*}
& \frac{\pi^2 R_c^2 r^4}{2} - \frac{\pi^2 R_c^4 r^2}{2} + \pi^2 R_c^2 r^2 - \frac{\pi^3 R_c^2 r^4}{2} - \frac{\pi^2 R_c^4}{2} - \frac{\pi^3 R_c^6}{2} \\
\sim &  \pi^2 R_c^2 r^2 - \frac{\pi^2 R_c^4}{2}, \quad r \rightarrow 0.
\end{align*}
}%
This completes the proof of \eqref{eq:ExpArea}.\hfill \IEEEQED

\bibliographystyle{IEEEtran}
\bibliography{MasterBibFile}

\end{document}

%% file: notation.tex
\def\nbo{{\mathbf{o}}}

\def\nbx{{\mathbf{x}}}
\def\nby{{\mathbf{y}}}

\def\nb0{{\mathbf{0}}}
\def\nb1{{\mathbf{1}}}

\def\nbbR{{\mathbb{R}}}

\newtheorem{lemma}{Lemma}

\newtheorem{prop}{Proposition}

\def\pc{\mathtt{P_c}}
\def\R{\mathbb{R}}

\def\sir{\mathtt{SIR}}

\def\se{\mathtt{SE}}

\newcommand{\dP}[1]{\mathbb{P}\left[#1\right]}
\newcommand{\dE}[2]{\mathbb{E}_{#2}\left[#1\right]}